\documentstyle[12pt]{article}
\begin{document}
\makeatletter
\@addtoreset{equation}{section} \renewcommand{\theequation}{\thesection.%
\arabic{equation}} \baselineskip=18.6pt plus 0.2pt minus 0.1pt

%%%%%%%%%%%%%%%%%%%%%%%%%%%%%%%%%%%%%%%%%%%%%%%%%%%%%%%%%%%%%%%%

\makeatletter \@addtoreset{equation}{section}
\begin{titlepage}
\title{
\hfill\parbox{4cm}{\normalsize UFR-HEP 00/11}\\
\hfill\parbox{4cm}{\normalsize LPHEA /00-04}\\ \vspace{2.5cm}
 Note on the Thermal Behavior of the Neutron Electric Dipole Moment from QCD Sum Rules }
\author{M. Chabab$^{1.2}\thanks{e-mail: mchabab@ucam.ac.ma}$,  N. El
Biaze$^1$ and R. Markazi$^1$ \\ \small{$^1$ Lab/UFR Physique des
Hautes Energies, Facult\'e des Sciences, B.P. 1014, Rabat,
Morocco.}\\ \small{$^{2}$ LPHEA, Physics  Department,
Faculty of Science- Semlalia, P.O. Box 2390, Marrakesh, Morocco.}}

\maketitle \thispagestyle{empty}

\begin{abstract}

We use the method of thermal QCD sum rules to investigate the
effects of temperature on the neutron electric dipole moment $d_n$
induced by the vacuum $\bar{\theta}$-angle. Then, we analyze and
discuss the thermal behaviour of the ratio $\mid {d_n \over
\bar{\theta}}\mid $ in connection with the restoration of the
CP-invariance at finite temperature.
\end{abstract}
\end{titlepage}
\newpage
\def\be{\begin{equation}}
\def\ee{\end{equation}}

\section{Introduction}
\qquad  The CP symmetry is, without doubt, one of the fundamental
symmetries in nature. Its breaking still carries a cloud of
mystery in particle physics and cosmology. Indeed, CP symmetry is
intimately related to theories  of interactions between elementary
particles and represents a cornerstone in constructing grand
unified and supersymmetric models. It is also necessary to explain
the matter-antimatter asymmetry observed in universe.

The first experimental evidence of CP violation was discovered in
the $K-\bar{K}$ mixing and kaon decays \cite{C}. According to the
CPT theorem, CP violation implies T violation. The latter is
tested through the measurement of the neutron electric dipole
moment (NEDM)$d_n$. The upper experimental limit  gives confidence
that the NEDM can be another manifestation of CP breaking. To
investigate the CP violation phenomenon many theoretical models
were proposed. In the standard model of electroweak interactions,
CP violation is parametrized by a single phase in the Cabbibo
Kobayashi Maskawa (CKM) quark mixing matrix \cite{CKM}. Other
models exhibiting a CP violation are given by extensions of the
standard model; among them, the minimal supersymmetric standard
model $MSSM$ includes in general soft complex parameters which
provide new additional sources of CP violation \cite{DMV,BU}.

 CP violation can be also investigated in the strong interactions context
through QCD framework. In fact, the QCD effective lagrangian
contains an additional CP-odd four dimensional operator embedded
in the following topological term:
\begin{equation}
L_{\theta}=\theta  {\alpha_s\over 8
\pi}G_{\mu\nu}\tilde{G}^{\mu\nu},
\end{equation}
where $G_{\mu\nu}$ is the gluonic field strength,
$\tilde{G}^{\mu\nu}$ is its dual and $\alpha_s$ is the strong
coupling constant. The  $G_{\mu\nu}\tilde{G}^{\mu\nu}$ quantity is
a total derivative, consequently it can contribute to the physical
observables only through non perturbative effects. The NEDM is
related to the $\bar \theta$-angle by the following relation :
\begin{equation}
d_n\sim {e\over M_n}({m_q\over M_n})\bar \theta \sim \{
\begin{array}{c}
2.7\times 10^{-16}\overline{\theta }\qquad \cite{Baluni}\\
5.2\times 10^{-16}\overline{\theta }\qquad \cite{cvvw}
\end{array}
\end{equation}
 and consequently, according to the experimental
measurements $d_n<1.1\times 10^{-25}ecm$ \cite{data}, the $\bar
\theta $ parameter must be less than $2\times10^{-10}$
\cite{peccei2}. The well known strong CP problem consists in
explaining the smallness of $\bar{\theta}$. In this regard,
several scenarios were suggested. The most known one was proposed
by Peccei and Quinn \cite{PQ} and consists in implementing an
extra $U_A(1)$ symmetry which permits a dynamical suppression of
the undesired $\theta $-term. This is possible due to the fact
that the axial current $J_5^\mu$  is related to the gluonic field
strength through the following relation $\partial_\mu
J_5^\mu={\alpha_s\over8 \pi}G_{\mu\nu}\tilde{G}^{\mu\nu}$. The
breakdown of the $U_A(1)$ symmetry gives arise to a very light
pseudogoldstone boson called axion. This particle may well be
important to the puzzle of dark matter and might constitute the
missing mass of the universe \cite{LS}.

Motivated by: (a) the direct relation between the $\bar
\theta$-angle and NEDM $d_n $, as it was demonstrated firstly in
\cite{cvvw} via the chiral perturbation theory and recently in
\cite{PR,PR1} within QCD sum rules formalism; (b) the possibility
to restore some broken symmetries by increasing the temperature;
we shall use the QCD sum rules at $T\ne 0$
 \cite{BS} to derive thermal dependence of the ratio $\mid{
d_n\over \bar{\theta}}\mid$. Then we study its thermal behaviour
at low temperatures and discuss the consequences of temperature
effects on the restoration of the broken CP symmetry.

 This paper is organized as follows: Section 2 is devoted to the calculations of the NEDM
induced by the  $\bar {\theta}$ parameter from QCD sum rules. In
section 3,  we show how one introduces temperature in QCD sum
rules calculations. We end this paper with a discussion and
qualitative analysis of the thermal effects on the CP symmetry.

\section{ NEDM from QCD sum rules}

\qquad In the two later decades, QCD sum rules \`a la SVZ
 \cite{SVZ} were
applied successfully to the investigation of hadronic properties
at low energies. In order to derive the NEDM through this
approach, many calculations were performed in the literature
\cite{CHM,KW}. One of them, which turns out to be more practical
for our study, has been obtained recently in \cite{PR,PR1}. It
consists in considering a lagrangian containing the following P
and CP violating operators:
\begin{equation}
L_{P,CP}=-\theta_q m_* \sum_f \bar{q}_f i\gamma_5 q_f +\theta
{\alpha_s\over 8 \pi}G_{\mu\nu}\tilde{G}^{\mu\nu}.
\end{equation}
$\theta_q$ and $\theta$ are respectively two angles coming from
the chiral and the topological terms and $m_*$ is the quark
reduced mass given by $m_*$=$m_um_d \over{m_u +m_d} $. The authors
of \cite{PR1} start from the two points correlation function in
QCD background with a nonvanishing $\theta$ and in the presence of
a constant external electomagnetic field $ F^{\mu\nu}$:
\begin{equation}
\Pi(q^2) = i \int d^4x
e^{iqx}<0|T\{\eta(x)\bar{\eta}(0)\}|0>_{\theta,F} .
\end{equation}
$\eta(x)$ is the interpolating current which in the case of the
neutron reads as \cite{I}:
\begin{equation}
\eta
=2\epsilon_{abc}\{(d^T_aC\gamma_5u_b)d_c+\beta(d^T_aCu_b)\gamma_5d_c\},
\end{equation}
where $\beta$ is a mixing parameter. Using the operator product
expansion (OPE), they have first performed the calculation of
$\Pi(q^2)$ as a function of matrix elements and Wilson
coefficients and then have confronted the QCD expression of
$\Pi(q^2)$ to its phenomenological parametrisation. $\Pi(q^2)$ can
be expanded in terms of the electromagnetic charge as\cite{CHM}:
\begin{equation}
\Pi(q^2)=\Pi^{(0)}(q^2)+e \Pi^{(1)}(q^2,F^{\mu\nu})+ O(e^2).
\end{equation}
The first term $\Pi^{(0)}(q^2)$ is the nucleon propagator which
include only the CP-even parameters \cite{SVZ1,IS}, while the
second term $\Pi^{(1)}(q^2,F^{\mu\nu})$ is the polarization tensor
which may be expanded through Wilson OPE as: $\sum
C_n<0|\bar{q}\Gamma q|0>_{\theta,F}$, where $\Gamma$ is an
arbitrary Lorentz structure and $C_n$ are the Wilson coefficient
functions calculable in  perturbation theory \cite{SVZ1}. From
this expansion, one keeps only the CP-odd contribution piece. By
considering the anomalous axial current, one obtains the following
$\theta$ dependence of $<0|\bar{q}\Gamma q|0>_{\theta}$ matrix
elements \cite{PR}:
\begin{equation}
m_q <0|\bar{q}\Gamma q|0>_{\theta}= i m_*\theta <0|\bar{q}\Gamma
q|0> ,
\end{equation}
where $m_q$ and $m_*$ are respectively the quark and reduced
masses. The electromagnetic dependence of these matrix elements
can be parametrized through the implementation of the $\kappa$,
$\chi $ and $\xi$ susceptibilities defined as \cite{IS}:\\
\begin{equation}
\begin{tabular}{lc}

$<0|\bar{q}\sigma^{\mu\nu} q|0>_F= \chi F^{\mu\nu} <0|\bar{q}q|0>$
\\ $g<0|\bar{q}G^{\mu\nu} q|0>_{F}= \kappa F^{\mu\nu}
<0|\bar{q}q|0> $ &   \\ $2g<0|\bar{q}\tilde{G}^{\mu\nu} q|0>_{F}=
\xi F^{\mu\nu} <0|\bar{q}q|0>.  $  &
\end{tabular}
\end{equation}

Putting altogether the above ingredients and after a
straightforward calculation \cite{PR1}, the following expression
of $\Pi^{(1)}(q^2,F^{\mu\nu})$
 for the neutron is derived:\\
\begin{eqnarray}
\Pi(-q^2)&=&-{\bar{\theta}m_* \over
{64\pi^2}}<0|\bar{q}q|0>\{\tilde{F}\sigma,\hat
q\}[\chi(\beta+1)^2(4e_d-e_u) \ln({\Lambda^2\over
-q^2})\nonumber\\ && -4(\beta-1)^2e_d(1+{1\over4}
(2\kappa+\xi))(\ln({-q^2\over \mu_{IR}^2})-1){1\over
-q^2}\nonumber\\ &&-{\xi\over
2}((4\beta^2-4\beta+2)e_d+(3\beta^2+2\beta+1)e_u){1\over
-q^2}...],
\end{eqnarray}
where  $\bar{\theta}=\theta+\theta_q$ is the physical phase and
$\hat q=q_\mu\gamma^\mu$. \\ The QCD expression (2.7) will be
confronted to the phenomenological parametrisation
$\Pi^{Phen}$$(-q^2)$ written in terms of the Neutron hadronic
properties. The latter is given by:\\
\begin{equation}
\Pi^{Phen}(-q^2)=\{\tilde{F}\sigma,\hat q\}
({\lambda^2d_nm_n\over(q^2-m_n^2)^2} +{A\over (q^2-m_n^2)}+...),
\end{equation}
where $m_n$ is the neutron mass, $e_q$ is the quark charge. A and
$\lambda^2$, which  originate from the phenomenological side of
the sum rule, represent respectively a constant of dimension 2 and
the neutron coupling constant to the interpolating current
$\eta(x)$. This coupling is defined via a spinor $v$ as
$<0|\eta(x)|n>=\lambda v$.

\section{ QCD sum rules at finite temperature}

\qquad The introduction of finite temperature effects may provide
more precision to the phenomenological values of hadronic
observables. Within the framework of QCD sum rules, the
T-evolution of the correlation functions appear as a thermal
average of the local operators in the Wilson
expansion\cite{BS,BC,M}. Hence, at nonzero temperature  and in the
approximation of the non interacting gas of bosons (pions), the
vacuum condensates can be written as :
\begin{equation}
<O^i>_T=<O^i>+\int{d^3p\over
2\epsilon(2\pi)^3}<\pi(p)|O^i|\pi(p)>n_B({\epsilon\over T})
\end{equation}
where $\epsilon=\sqrt{p^2+m^2_\pi}$, $n_B={1\over{e^x-1}} $is the
Bose-Einstein distribution and $<O^i>$ is the standard vacuum
condensate (i.e. at T=0). In the low temperature region, the effects of heavier resonances 
$(\Gamma= K, \eta,.. etc)$ can be neglected due to their distibution functions $\sim e^{-
m_\Gamma \over T}$\cite{K}. To compute the pion matrix elements, we
apply the soft pion theorem given by:
\begin{equation}
<\pi(p)|O^i|\pi(p)>=-{1\over f^2_\pi}<0|[F^a_5,[F^a_5,O^i]]|0>+
O({m^2_\pi \over \Lambda^2}),
\end{equation}

where $ \Lambda$ is a hadron scale and $F^a_5$ is the isovector
axial charge:
\begin{equation}
F^a_5=\int d^3x \bar{q}(x)\gamma_0\gamma_5{\tau^a\over2}q(x).
\end{equation}
Direct application of the above formula to the quark  and gluon
condensates shows that \cite{GL,K}:\\ (i) Only $<\bar{q}q>$ is
sensitive to temperature. Its behaviour at finite T is given by:
\begin{equation}
<\bar{q}q>_T\simeq (1-{\varphi(T)\over8})<\bar{q}q>,
\end{equation}
where $\varphi(T)={T^2\over f^2_\pi}B({m_\pi\over T})$ with $B(z)=
{6\over\pi^2}\int_z^\infty dy {\sqrt{y^2-z^2}\over{e^y-1}}$ and
$f_\pi$ is the pion decay constant ($f_\pi\simeq 93 MeV$). The
variation with temperature of the quark condensate $<\bar{q}q>_T$
results in two different asymptotic behaviours, namely: \\
$<\bar{q}q>_T\simeq (1-{T^2\over {8f^2_\pi}})<\bar{q}q>$ \quad for
 ${m_\pi\over T}\ll 1$, and $<\bar{q}q>_T\simeq (1-{T^2\over
{8f^2_\pi}}e^{-m_\pi \over T})<\bar{q}q>$ \quad for ${m_\pi\over
T}\gg 1$.\\ (ii) The gluon condensate is nearly constant at low
temperature and a T dependence occurs only at order $T^8$.

As usual, the determination of the ratio ${d_n \over
\bar{\theta}}$ sum rules at non zero temperature  is now easily
performed through two steps. In the first step, we apply Borel
operator to both expressions of the Neutron correlation function
shown in Eqs. (2.7) and (2.8), where finite temperature effects
were introduced as discussed above. Next step, by invoking the
quark-hadron duality principle, we deduce the following relation
of the $\bar { \theta}$ induced NEDM:
\begin{equation}
{d_n\over \bar{\theta}}(T)=-{M^2m_* \over 16\pi^2}{1\over
\lambda_n^2(T)M_n(T)}(1-{\varphi(T)\over
8})<\bar{q}q>[4\chi(4e_u-e_d)-{\xi\over 2M^2}(4e_u+8e_d)]e^{M_n^2
\over M^2},
\end{equation}
where M represents the Borel parameter. Note that we have
neglected the single pole contribution entering via the constant
A, as suggested in \cite{PR}.\\ The expression (3.5) is derived
with $\beta=1$ which is more appropriate for us since it
suppresses the infrared divergences. In fact, the Ioffe choice
$(\beta=-1)$ which is rather more useful for the CP even case,
removes the leading order contribution in the sum rules (2.7). The
coupling constant $\lambda_n^2(T)$ and the neutron mass $M_n(T)$
which appear in (3.5) were determined from the thermal QCD sum
rules. For the former, we consider the $\hat q $ sum rules in
\cite{K,J} with $\beta=1$ and  then we extract the following
explicit expression of $\lambda_n^2(T)$:
\begin{equation}
\lambda_n^2(T)=\{{3\over{8{(2\pi)}^4}}M^6+{3\over{16{(2\pi)}^2}}M^2<{\alpha
_s\over\pi} G^2>\}\{1-(1+{g^2_{\pi NN}f^2_\pi\over
M_n^2}){\varphi(T) \over 16}\}e^{M^2_n(0)\over M^2}
\end{equation}

Within the pion gas approximation, Eletsky has demonstrated in
\cite{E} that inclusion of the contribution coming from the
pion-nucleon scattering in the nucleon sum rules is mandatory. The
latter enters Eq.(3.6) through the coupling constant $g_{\pi NN}$,
whose values lie within the range 13.5-14.3 \cite{PROC}.
\\ \qquad Numerical analysis is performed with the following input
parameters: the Borel mass has been chosen within the values
$M^2=0.55-0.7GeV^2$ which correspond to the optimal range (Borel
window) in the $ d_n\over \bar{\theta}$ sum rule at $T=0$
\cite{PR1}. For the $\chi$ and $\xi$ susceptibilities we take
$\chi=-5.7\pm 0.6 GeV^{-2}$ \cite{BK} and $\xi=-0.74\pm 0.2$
\cite{KW}. As to the vacuum condensates appearing in (3.5), we fix
$<\bar{q}q>$ and $<G^2>$ to their standard values \cite{SVZ}.

\section{Discussion and Conclusion}

\qquad In the two above sections, we have established the relation
between the NEDM and $ \bar{\theta} $ angle at non zero
temperature from QCD sum rules. Since the ratio  ${ d_ n \over
\bar{\theta}}$ is expressed in terms of the pion parameters
$f_\pi$, $m_\pi$ and of $g_{\pi NN}$, we briefly recall the main
features of their thermal behaviour.
 Various studies performed either
within the framework of the chiral perturbation theory and/or QCD
sum rules at low temperature have shown the following features:
\\ (i) The existence of QCD phase
transition temperature $T_c$ which signals both QCD deconfinement
and chiral symmetry restoration \cite{BS,G}.\\ (ii) $f_\pi$ and
$g_{\pi NN}$ have very small variation with temperature up to
$T_c$. So, we shall assume them as constants below $T_c$. However,
they vanish if the temperature passes through the critical value
$T_c$ \cite{DVL}.\\ (iii) The thermal mass shift of the neutron
and the pion is absent at order $O(T^2)$ \cite{K,BC}. $\delta M_n$
shows up only at the next order $T^4$, but its value is negligible
\cite{E}.\\ By taking into account the above properties, we plot
the ratio defined in Eq. (3.5) as a function of T. From the
figure, we learn that  the ratio $\mid{ d_ n \over \bar{\theta}
}\mid$ survives at finite temperature and it decreases smoothly
with T (about 16$\%$ variation for temperature values up to 200
MeV). This means that either the NEDM value decreases or
$\bar{\theta}$ increases. Consequently, for a fixed value of
$\bar{\theta}$ the NEDM decreases but it does not exhibit any
critical behaviour. Furthermore, if we start from a non vanishing
$ \bar{\theta} $ value at $T=0$, it is not possible to remove it
at finite temperature. We also note that $ \mid{d_n\over
\bar{\theta}}\mid$ grows as
 $M^2$ or $\chi$ susceptibility increases.
It also grows with quark condensate rising. However this ratio is
insensitive to both the $\xi$ susceptibility and the coupling
constant $g_{\pi NN}$. We notice that for higher temperatures, the
curve $\mid{d_n\over \bar{\theta}}\mid=f({T\over T_c})$ exhibits a
brutal increase justified by the fact that for temperatures beyond
 the critical value $T_c$, at which the chiral symmetry is restored,
the constants $f_\pi$ and $g_{\pi NN}$ become zero and
consequently from  Eq(3.5) the ratio $ {d_n\over \bar{\theta}}$
behaves as a non vanishing constant. The large difference between
the values of the ratio for $T<T_c$ and $T>T_c$ maybe a consequence
of the fact that we have neglected other  contributions
to the the spectral function, like the scattering process $ N+ \pi \to \Delta $. These contributions
, which are of the order $T^4$, are negligible in the low temperture region but become 
substantial for $T\ge T_c$.
Moreover, this difference may also originate from the use of soft
pion approximation which is valid essentially for low $T$ ($T<
T_c$). Therefore it is clear from this qualitative 
analysis, which is based on the soft pion approximation, that the
temperature does not play a fundamental role in the suppression of
the undesired $\theta$-term and hence the broken CP symmetry is
not restored, as expected. This is not strange, in fact it was
shown that more heat does not imply automatically more symmetry
\cite{MS,DMS}. Moreover, some exact symmetries can be broken by
increasing temperature \cite{W,MS}. The symmetry non restoration
phenomenon, which means that a broken symmetry at T=0 remains
broken even at high temperature, is essential for discrete
symmetries, CP symmetry in particular. Indeed, the symmetry non
restoration allows us to avoid wall domains inherited after the
phase transition \cite{ZKO} and to explain the baryogenesis
phenomenon in cosmology \cite{S}. Furthermore, it can be very
useful for solving the monopole problem in grand unified theories
\cite{DMS}.

{\bf Acknowledgments}

We are deeply grateful to T. Lhallabi and E. H. Saidi for their
encouragements and stimulating remarks. N. B. would like to thank
the Abdus Salam ICTP for hospitality and Prof. Goran Senjanovic
for very useful discussions.\\ This work is supported by the
program PARS-PHYS 27.372/98 CNR and the convention CNPRST/ICCTI
340.00.

\newpage
\section*{\bf Figure Captions}
Figure: Temperature dependence of the ratio $\mid{d_n \over
\bar{\theta}}\mid$
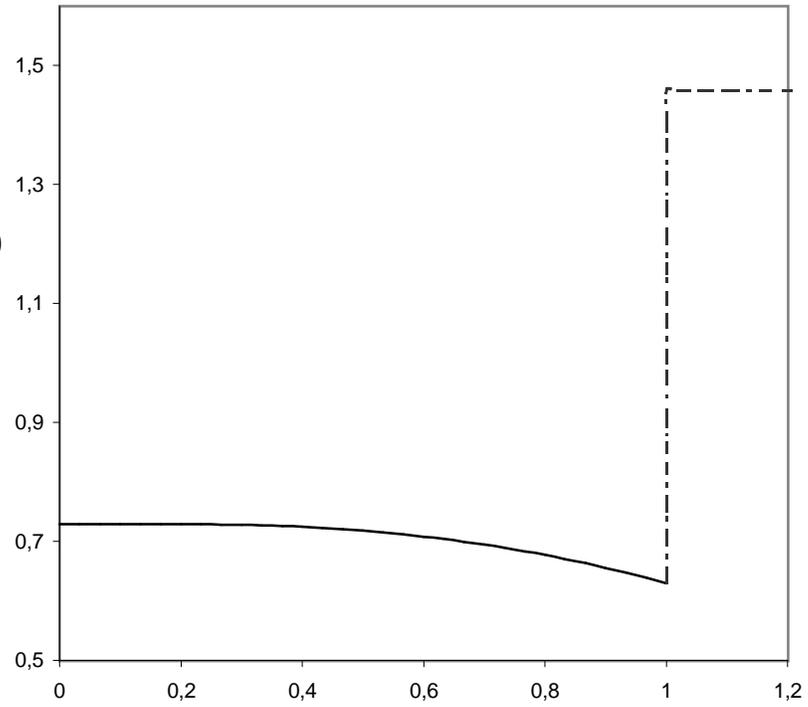
\end{document}